\documentclass[conference]{IEEEtran}

\usepackage{makeidx}  
\usepackage{times,amsmath,epsfig}
\usepackage{epstopdf}
\usepackage{amsmath}
\usepackage{listings}  
\usepackage{color}
\usepackage{rotating}
\usepackage{enumerate}
\usepackage{url}
\usepackage{multirow}
\usepackage{tabu}
\usepackage{amssymb}
\usepackage{graphicx}
\usepackage{float}
\usepackage{hyperref}
\usepackage[utf8]{inputenc}
\usepackage{footnote}
\usepackage{todonotes}

\begin{document}
%

\title{Towards Cleaning-up Open Data Portals: \\ A Metadata Reconciliation Approach}

%
\author{
\IEEEauthorblockN{Alan Tygel+, S\"{o}ren Auer*, Jeremy Debattista*, Fabrizio Orlandi*, Maria Luiza Machado Campos+}\\
\IEEEauthorblockA{+ Graduate Program on Informatics -- PPGI -- UFRJ, Brazil\\
\{alantygel, mluiza\}@ppgi.ufrj.br
\\
* University of Bonn and Fraunhofer IAIS, Germany\\
\{auer, debattis, orlandi\}@cs.uni-bonn.de}
}

\maketitle
\IEEEpeerreviewmaketitle

\begin{abstract}

This paper presents an approach for metadata reconciliation, curation and linking for Open Governamental Data Portals (ODPs).
ODPs have been lately the standard solution for governments willing to put their public data available for the society.
Portal managers use several types of metadata to organize the datasets, one of the most important ones being the tags.
However, the tagging process is subject to many problems, such as synonyms, ambiguity or incoherence, among others.
As our empiric analysis of ODPs shows, these issues are currently prevalent in most ODPs and effectively hinders the reuse of Open Data.
In order to address these problems, we develop and implement an approach for tag reconciliation in Open Data Portals, encompassing local actions related to individual portals, and global actions for adding a semantic metadata layer above individual portals.
The local part aims to enhance the quality of tags in a single portal, and the global part is meant to interlink ODPs by establishing relations between tags.

\end{abstract}


\section{Introduction}
\label{sec:introduction} 

Analysing large amounts of data plays an increasingly important role in today's society. 
However, new discoveries and insights can only be attained by integrating information from dispersed sources. 

One approach for addressing the problem of data dispersion are data catalogues, which enable organizations to upload and describe datasets using comprehensive metadata schemes. 
Similar to digital libraries, networks of such catalogues can support the description, archiving and discovery of datasets on the Web. 
Recently, we have seen a rapid growth of data catalogues being made available to the public. 
The data catalogue registry \href{http://datacatalogs.org}{datacatalogs.org}, for example, already lists 285 data catalogues worldwide. 

Data catalogues where data is supposed to be open, at least in the licensing sense, are usually called Open Data Portals (ODPs).
Implementations that show the increasing popularity of ODPs can be seen, for example, in open government data portals, data portals of international organizations and NGOs, as well as scientific data portals.

These ODPs comprise large amounts of structured data, mostly in the form of tabular data such as CSV files or Excel sheets. 
They aim to be a one-stop-shop for citizens and companies interested in using public data produced by governments or civil society organisations.
Examples are the \href{http://data.gov}{US' data portal}, the \href{http://data.gov.uk}{UK's data portal}, the \href{http://open-data.europa.eu}{European Commission's} portal as well as numerous other local, regional and national data portal initiatives.

In the research domain ODPs also play an important role.
An example of a popular scientific open data portals is the \href{http://data.gbif.org}{Global Biodiversity Information Facility Data Portal}.
Also many international and non-governmental organizations operate ODPs such as the \href{http://data.worldbank.org}{World Bank Data Portal} or the data portal of the \href{http://apps.who.int/gho/data/}{World Health Organization}.

Despite its recent popularity, Open Data and Open Data Portals still face significant impediments, as richly described in~\cite{Zuiderwijk2012}.
Zuiderwijk et al. collected 118 socio-technical impediments for use of open data from interviews, workshops and literature.
Some cited impediments were ``absence of commonly agreed metadata'', ``insufficiency of metadata'', ``the lack of interoperability'' and ``difficulty in searching and browsing data'', showing that a great challenge for ODPs is the organization of data.

The open data organization challenge can be subdivided into two aspects: 1) structuring and organizing the datasets themselves and 2) providing well-structured and organized metadata for the datasets.
The first aspect was, for example, tackled by approaches for semantic lifting of data by~\cite{DBLP:conf/i-semantics/ErmilovAS13} and~\cite{Ding2011325}, who tried to build general strategies for putting large open government datasets in the Link Data cloud.
For the standardized structuring metadata, the Data Catalog Vocabulary (DCAT)\footnote{Available at \url{http://www.w3.org/TR/vocab-dcat/}}~\cite{conf/i-semantics/CyganiakMP10} was developed.
However, the cross-portal metadata alignment and reconciliation can not be addressed by DCAT.

The metadata used to organize datasets in an ODP comprises categories or groups and most importantly labelling with free-text words or sets of words -- the tags.
The concept of tagging became popular within Web 2.0 services and aggregation tools like del.icio.us. 
The main advantages of tagging are the ease of classifying, and the crowd effect -- resulting in the so called folksonomies -- because all users were allowed to tag and share their contents. 
Tagging datasets in an ODP cannot be considered as folksonomies, because the process is mainly driven by portal managers and data publishers, and not by the actual users.
As a result of this, the structuring effect of crowd-tagging and folksonomies is missing in ODPs.

A quick look over some ODPs reveals that most of them suffer from a very confusing organization of datasets. 
The first level of categorization uses the concept of groups.
In general, they are stable and meaningful, but normally contain a large number of datasets.
A more detailed classification should be done via tags, whose use in ODPs has the following issues:

\begin{itemize}
	\item \emph{Synonyms:} In most ODPs, there exists large number of synonymous tags;
	\item \emph{Misspellings:} Several tags are incorrectly written, and have differences in capitalization or accents;
	\item \emph{Lack of relationships:} There is no explicit relationships between the tags;
	\item \emph{Ambiguity:} As tags are written as pure text, ambiguity is prevalent in ODPs; and
	\item \emph{Incoherence:} Tags do not allow any connection between different portals that use the same or equivalent tags.
\end{itemize}

As a result, the navigation, exploration and search within individual, but in particular also across ODPs is significantly hampered.

In this paper, we present an approach and its implementation for improving the tag curation within and across ODPs.
Our main contributions are:
\begin{itemize}
	\item A comprehensive analysis of tag usage in 90 ODPs, which justifies the need and benefits of better tools for managing tags;
	\item An approach for cleaning and reconciliation of tags in ODPs; and
	\item An approach for collaboratively connecting ODPs through meaningful shared tags.
\end{itemize}

The remainder of this paper is organized as follows.
After deriving a basic characterization, we present an analysis of the use tags in ODPs in \autoref{sec:analysis}.
\autoref{sec:stodap} shows our approach, both at the local (individual ODPs) and global levels (Tag Server).
\autoref{sec:implementation} describes the implementation of the open source tools, while \autoref{sec:results} describes some of the achieved results.
The final sections outline related work and draw the conclusions of the paper.

\section{Characterization of an Open Data Portal}
\label{sec:characterization} 

According to~\cite{Colpaert2013}, an Open Data Portal is ``a collection of systems set up to make Open Data used and useful''.
A formal definition of an ODP can be found in~\cite{Umbrich2015}.
However, in that case, the focus is general metadata analysis, which turns their definition unsuitable to be used here. 

\begin{figure}
\begin{center}
\includegraphics[width=\columnwidth]{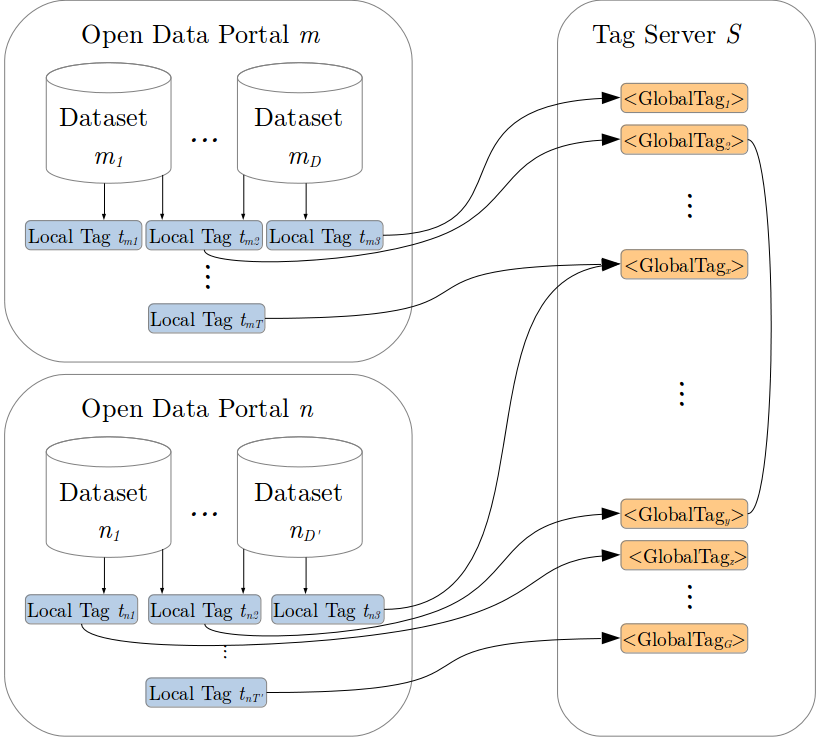}
\caption{Relevant elements of the Semantic Tags for Open Data Portals system.}
\label{fig:definition}
\end{center}
\end{figure}

\autoref{fig:definition} shows the relevant entities and relations that are used in the remainder of this paper.
An \emph{Open Data Portal}, in this context, is a collection of datasets, which hold open data resources online.
Each \emph{Dataset} belonging to an ODP can be tagged with local tags. 
Each \emph{Local Tag} also belongs to an ODP, and can be used to tag one or more datasets.
In this architecture, local tags are connected to \emph{Global Tags}, stored in a collaborative \emph{Tag Server}.
Several local tags from different ODPs can be associated to a single global tag, which can also have semantic relationships with other global tags.

\section{Analysis of The Use of Tags in Open Data Portals}
\label{sec:analysis} 

In this section, we profile the use of tags in Open Data Portals. 
The analysis is restricted to systems running CKAN\footnote{Available at \url{http://ckan.org}}, the standard open-source software for ODPs. 
The CKAN community publishes a census\footnote{Available at \url{http://ckan.org/instances}}, were 139 portals are listed. 
Through the API offered by the software, we tried to obtain data from all portals, but only 90 responded adequately when the assessment was performed (Sep. 2015).
Reasons for the lack of availability were mainly that the portal was completely offline, the API was disabled or not responding at the same URL of the website or the portal was using an outdated version of CKAN. 

Most of the ODPs are related to governments and public administrations at local, regional, national or continental levels. 
Some of them are also focusing on specific themes, such as energy or geothermal data. 
Although most ODPs are authoritative and run by governments and public administrations, some of the portals were built as civil society initiatives.
A complete list of the analysed ODPs is available online\footnote{\url{http://bit.ly/1NGygtk}}.

The analysed ODPs are quite heterogeneous. 
The number of datasets in each portals varies from 3 to 147,485, and the number of tags, from 3 to 49,189.
Regarding the quality of the portals, although there is no general benchmark, \emph{\href{http://opendatamonitor.eu}{Open Data Monitor}} attests a high heterogeneity within European ODPs.
An informal quality assessment using the Five Stars of ODPs~\cite{Colpaert2013} also shows that portals vary from simple data registries (one star) to a common data hub (five stars).

A summary of the experiment data is shown in \autoref{tab:summary}. 
The code used to collect and analyse the data is available as an open-source project\footnote{\url{https://github.com/alantygel/StodAp}}.

\begin{savenotes}
\begin{table}[tb]
\caption{Summary of data used in the experiment.}
\label{tab:summary}
\begin{center}
\begin{tabular}{l|c}
Portals & 90 \\
Datasets & 	389,913 \\
Overall tags & 220,567 \\
Unique tags & 148,657 \\
Average tags per portal & 2451 \\	
Average tags per dataset & 3.88 \\	
Association with semantic resources & 36\% \\
Groups & 1500 \\
Average groups per portal\footnote{Excluding ODPs which do not use groups.} & 21.43 \\
Average datasets per group\footnote{Excluding void groups.} & 67.45 \\
\end{tabular}
\end{center}
\end{table}
\end{savenotes}

The analysis is divided in two groups: local metrics, to analyse the quality of tags in a particular ODP, and global metrics, looking at the interrelations between portals, and with the Linked Open Data (LOD) cloud.

Regarding the other main tool for organizing ODPs -- groups -- \autoref{tab:summary} also shows the number of groups per portal, and the number of datasets inside each one.
While the tags are attributed to an average 3.88 datasets, groups contain a mean value of 67.45 datasets.
This makes groups less selective than tags, which justifies our decision to focus on tags in this work.
Moreover, while all 90 portals use tags, 20 do not use groups to organize data.

\subsection{Local Metrics}
\label{sec:local_metrics}

\subsubsection{Tag Reuse}
\begin{figure}[tb]
\begin{center}
\includegraphics[width=\columnwidth]{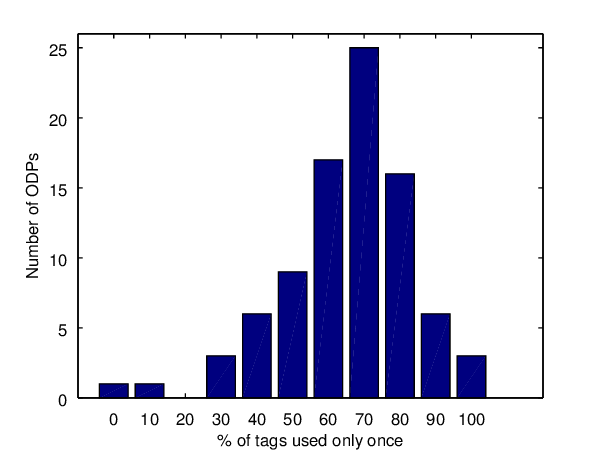}
\caption{Re-use of tags inside a portal. The graphic shows the distribution of the percentage of tags used only once.}
\label{fig:tags_once}
\end{center}
\end{figure}

The objective of this metric is to assess whether a single tag is being used to characterize several datasets, just a few or even only one.
Creating new tags for each dataset can be considered a bad tagging practice.
If tags are reused for several datasets, tag-based information retrieval will be more effective.
\autoref{fig:tags_once} shows the distribution of the percentage of tags used only once for each portal. 
The graphic shows a peak around 70\% of the tags used only once.
From the 90 portals, 78 use more than 50\% of the tags only once.
As a conclusion, tag reuse can be considered very low, thus effectively preventing the tags to be a suitable means to improve navigation, exploration and retrieval of datasets from ODPs.


\subsubsection{Tags per dataset}
\begin{figure}[tb]
\begin{center}
\includegraphics[width=\columnwidth]{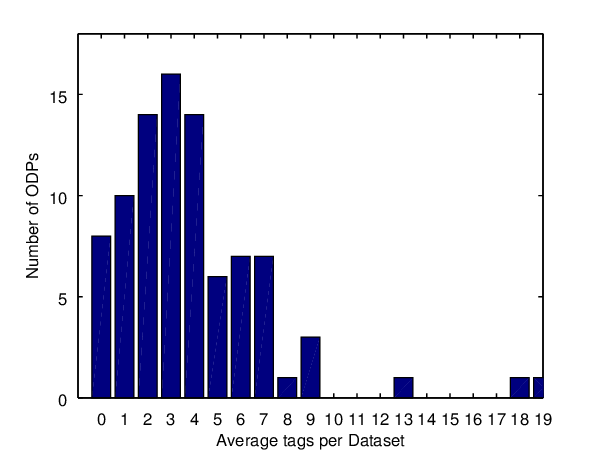}
\caption{Distribution of the average number of tags used per dataset in Open Data Portals.}
\label{fig:tags_per_ds}
\end{center}
\end{figure}

This metric assesses the number of tags used per dataset.
The goal is to verify, as in~\cite{Umbrich2015}, if the tag metadata is being actively used in the portals.
However, the results of this metric cannot lead to further conclusions, since there is no optimal value for the number of tags per dataset. 
Using few and consistently used tags may support the organization of datasets better than many incoherently used ones.
On the other hand, few tags may not label the content adequately.
\autoref{fig:tags_per_ds} shows the distribution of the average tags per dataset for each portal. 
We can see that most ODPs apply between 1 and 7 tags to each dataset, with a peak around the value of 3.
In general, we can affirm that describing datasets with tags is a common procedure in ODPs.

\subsubsection{Tag similarity}
By looking at the ODP tags, one can readily recognize that many tags differ only on capitalization, accents or singular and plural forms.
Thus, this metric assesses whether several tags are being used with the same meaning.
While recognizing these cases is easy for humans who understand the language of the tags, an automatic discovery of tags with the same meaning is not always straightforward.
A simple approach is to convert the tags to lowercase and unaccented strings for comparison. 
Despite its simplicity, this method catches a significant number of cases such as \texttt{birth} and \texttt{Birth}.

A second possibility is to use the well known Levenshtein edit distance, which can also be suitable for detecting gender and plural differences, in some languages. 
However, this method fails with tags containing numbers. 
For example, the Levenshtein edit distance between \texttt{budget-2010} and \texttt{budget-2011} is the same as between \texttt{Access} and \texttt{access}.
Semantic-oriented methods, as detailed in~\cite{Harispe2015}, could also be used to detect synonymous tags.

\begin{figure}[tb]
\begin{center}
\includegraphics[width=\columnwidth]{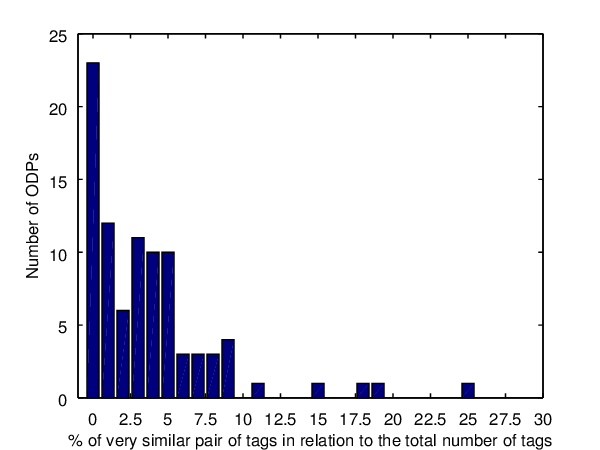}
\caption{Proportion of similar tags in ODPs, where the difference lies only capitalization or special characters.}
\label{fig:similarity}
\end{center}
\end{figure}

\autoref{fig:similarity} shows a distribution of the percentage of similiar tags inside each ODP.
Similarity was checked using the simple approach.
The occurrence of a significant rate of similarity reveals that there are few portals adopting a systematic tagging procedure.
Despite the low percentage for some portals, in many of them similar tags still occur.
Only 20 portals, out of overall 90, revealed no similar tags at all. 
It should be noticed that these portals use far less tags (148 per portal) than the average of all portals (2451 per portal), which may also be a sign of careful tagging.

\subsection{Global Metrics}

\subsubsection{Coincident tags between portals}
Different ODPs, especially governmental ones, can publish related data, which may also be tagged similarly. 
Using the same tag comparison approach as described in the local tag similarity metric, we found that 73,316 tags appeared in more than one ODP, which represents 33\% of the total tags. 
If we are interested in datasets from different ODPs tagged similarly, an overestimation bias may come from the fact that some portals act only as datasets harvesters, replicating the same datasets (and related tags). 
On the other hand, because portals are available in several languages, different tags could have the same meaning in different languages, what in turn tends to be an underestimation bias.
In any case, the figure clearly indicates that there exists great potential for linking tags between open data portals.
In fact, with this metric, our aim is to justify and motivate the development of a semantic tag curation approach for open data portals, which will be described in \ref{sec:global}. 

\subsubsection{Tag expressiveness}
A way of taking the tagging process one step further is to associate tags with resources or terms described in knowledge bases.
In~\cite{Passant2008}, while building the MOAT ontology\footnote{\url{http://muto.socialtagging.org/mirror/moat.rdf}}, the authors designed the association of each tag with a meaning, represented by one or more URIs in the LOD cloud.
With the expressiveness metric, our aim is to check if a tag is suitable to be connected to LOD cloud, i.e., if there are possible resources to represent its meaning.

In order to search for candidate resources for the tags, we used \url{Lexvo.org}~\cite{Melo2013}, a service that offers connections to different semantic knowledge bases, in several languages.
By providing a term (in our case, the tag) and its language, Lexvoc.org returns the corresponding resources, either as \texttt{rdfs:seeAlso} or \texttt{lexvo:means}.

\autoref{tab:expressiveness} shows the results. 
The majority of tags (68.38\%) did not correspond to any semantic resource according to this method.
8.15\% of the tags were not evaluated either because they contain numbers, or because their length was equal or smaller than three. 
In those cases, results are mostly wrong.
For 23.46\% of the tags, at least one meaning or equivalent term was found, and their use represent a similar magnitude of 23.71\%.

\begin{table}[]
\centering
\caption{Expressiveness of tags}
\label{tab:expressiveness}
\begin{tabular}{|l|c|c|}
\hline
                            & \textbf{Absolute Occurrence} & \textbf{Weighted by Usage} \\ \hline
Associated to a meaning     & 23.46\%           & 23.71\%                    \\ \hline
Not associated to a meaning & 68.38\%           & 64.20\%                    \\ \hline
Not considered              & 8.16\%            & 12.09\%                    \\ \hline
\end{tabular}
\end{table}

It is not possible to guarantee that all associations were meaningful, and even worse, that the meaning intended by the tagger was correctly captured.
The tag language was estimated by the ODP locale, which can also be a source of errors if not correctly set.
Further evaluations are needed in order to estimate the potential that ODP tags have to be connected to the LOD cloud.
However, we see that at least one fifth of the tags correspond directly to a semantic resource.
Providing context and a stemming pre-processing would probably enhance this result.
Thus, we can say that some semantic potential is present on the tags.

\hspace{0.2pt}

After this analysis, we can affirm that: (i) tags in ODPs are widelly used, but in a non-systematic way, which hinders their capacity of supporting information retrieval, and (ii) there is a potential for using these tags as connecting elements between ODPs, and for raising semantics from them.
Next, we describe our proposal based on these statements.





\section{Semantic Tags for Open DAta Portals - The STODaP Approach}
\label{sec:stodap}

In this section, we describe a tag reconciliation approach for cleaning up ODPs, supported by software tools both at the local and global contexts.
The objective is to tackle the main problems identified by the metrics described in the previous section, and thus to facilitate data organization and linking through metadata descriptions of ODPs.

\begin{figure*}[tb]
\begin{center}
\includegraphics[width=\textwidth]{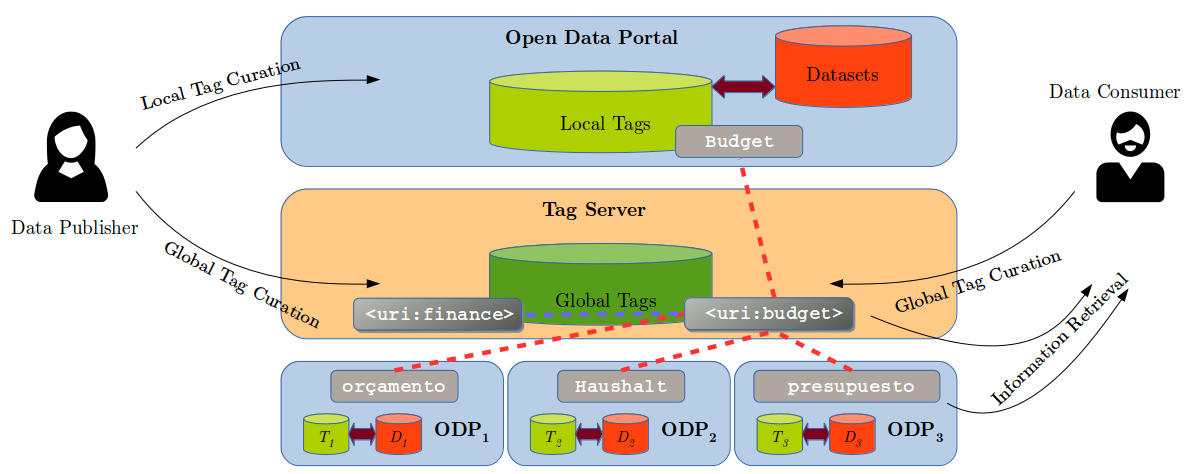}
\caption{Overview of the StodAp approach. Local tags are connected to a corresponding global tag within a central tag server. 
Data managers responsible for an ODPs may use tools for local tag curation, as well for maintaining the tag server. 
This task is also expected to be performed by data consumers.}
\label{fig:overview}
\end{center}
\end{figure*}

\autoref{fig:overview} shows an overview of the proposed approach\footnote{Icons by \href{http://www.flaticon.com/authors/simpleicon}{SimpleIcon} from \href{http://www.flaticon.com}{www.flaticon.com} are licensed under \href{http://creativecommons.org/licenses/by/3.0/}{CC BY 3.0}.}.
Data publishers in charge of ODPs are offered tools for local tag curation.
These tags are then connected to global tags hosted in a central Tag Server, which can be collaboratively edited both by data consumers and publishers. 
They can add new semantic descriptions to the global tags, establish relations between them, and also create new links between global and local tags.
Data consumers have the option to retrieve data directly from ODPs, or through references gathered from the central server.
The description of these actions is shown in the sequel.

\subsection{Local Approach: cleaning up tags inside an ODP}
\label{sec:local}

\autoref{sec:local_metrics} showed that ODPs suffer from low reuse of tags, and that there is a significant tags duplication due to slight spelling differences. 
In fact, both problems -- low reuse and duplication -- are connected, since merging similar tags improves tag reuse.
However, low tag reuse can be also attributed to the lack absence of a standard tagging procedure, which would guide users in this task.

To address this problem locally at a particular ODP, we propose an approach for reconciliation of tags. 
First, we offer three levels of semi-automatic tag merging strategies:



\begin{enumerate}
	\item With high confidence, we suggest merging tags that differ only by capital letters or special characters. 
In many ODPs, this strategy will already achieve significant results, as shown in~\autoref{fig:similarity}.
	\item After running the first strategy, the Levenshtein distance is computed for all remaining pairs of tags.
Tags with distance one or two are suggested for merging, in order to catch plural/gender variations, such as \texttt{worker} and \texttt{workers}. 
However, false-positives like \texttt{widow} and \texttt{window} may appear.
Tags containing numbers (to avoid merging tags containing years) or less than 4 characters are not included.
	\item Finally, we use semantic measures~\cite{Harispe01032014} to determine the semantic similarity between two tags. 
In this case, the tags \texttt{autumn} and \texttt{fall} have a high similarity, and thus will be suggested for merging.
\end{enumerate}

After this cleaning procedure, we offer users the opportunity to link each local tag to a global correspondent at the tag server, described in the sequel.

\subsection{Global Approach: a model for linking tags between ODPs}
\label{sec:global}

With the aim of building a common and collaborative basis for interlinking ODPs, we developed a Global Tag Server.
The conceptual rationale is:
\begin{enumerate}
	\item To assist individual ODPs enhancing the quality of their tags, by assigning a common agreed meaning to them;
	\item To create a collaborative platform for meaningfully linking ODPs.
\end{enumerate}

\begin{figure}[tb]
\begin{center}
\includegraphics[width=\columnwidth]{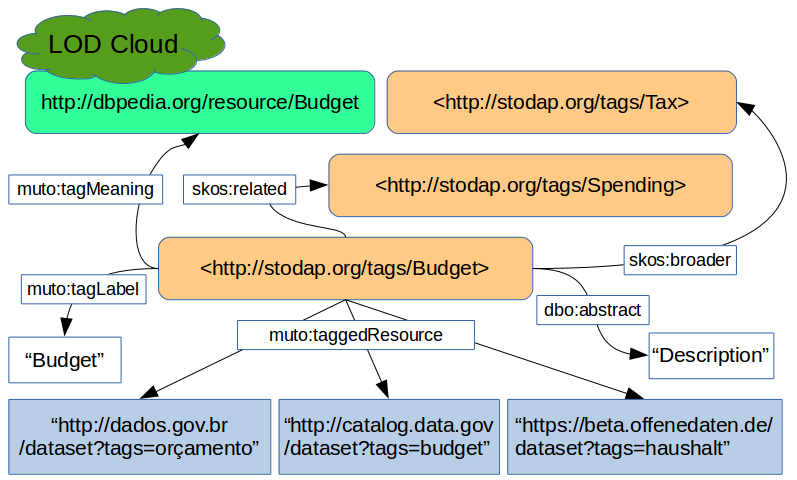}
\caption{Example of the StodAp model showing relationships of the global tag \url{http://stodap.org/tags/Budget}.}
\label{fig:example}
\end{center}
\end{figure}

The Global Tag Server hosts the description of global tags.
Each global tag may be associated to one or more Linked Open Data resources, representing their semantic meanings.
Linking to the local tags is accomplished via the URIs which represent a local tag in its context.
The global tags can also have several types of relations between each other, such as \texttt{skos:broader}, \texttt{skos:narrower} or \texttt{owl:sameAs}.
\autoref{fig:example} illustrates the concept with an example.

The example shows the global tag identified by the URI \texttt{<http://stodap.org/tags/Budget>}. 
With this global tag, a meaning and some URIs of local tags are associated.
The global tag is also semantically related to other global tags, using the SKOS vocabulary.
The MUTO ontology\footnote{\url{http://muto.socialtagging.org/core/v1.html}} is used to define some concepts and relations between the tags, like \texttt{muto:Tag}, \texttt{muto:taggedResource}, \texttt{muto:hasTag} and \texttt{muto:hasMeaning}.



\section{Implementation}
\label{sec:implementation}

In order to support the StodAp approach, we describe in this section some software tools that were implemented.
For the local tag curation, we implemented two CKAN plugins: 
(i) \emph{CKAN Tag Manager}\footnote{\url{https://github.com/alantygel/ckanext-tagmanager}} and 
(ii) \emph{CKAN Semantic Tags}\footnote{\url{https://github.com/alantygel/ckanext-semantictags}}.

\begin{figure}[tb]
\begin{center}
\includegraphics[width=\columnwidth]{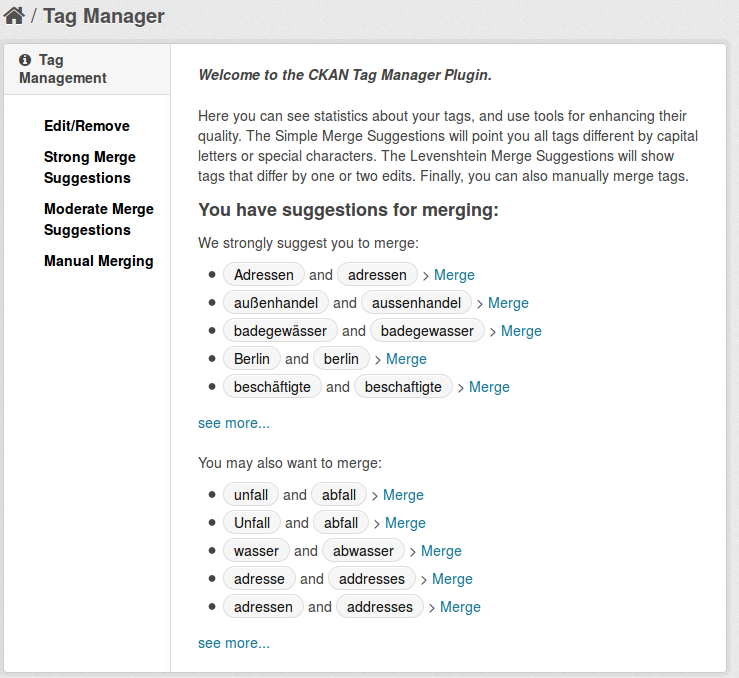}
\caption{Local tag curation in a CKAN instance. The plugin offers possibilities of manual and semi-automatic tag merging. The first block contains only valid suggestions, while the second block shows 3 false-positives. The tags in this example were extracted from the German offenedaten.de portal. }
\label{fig:local_curation}
\end{center}
\end{figure}

The CKAN Tag Manager one offers an environment for tag curation directly inside the CKAN platform. 
It comprises basic functions such as deletion and editing of tags, and advanced function aimed to enhance the quality of tags.
In this sense, the plugin checks:
\begin{itemize}
	\item Very similar tags, differing by capitals or special characters;
	\item Similar tags, with a Levenshtein distance $\le 2$ (after lowercasing and unaccenting)
\end{itemize}
In all those cases, the user is offered the option of merging the respective pair of tags. 
\autoref{fig:local_curation} shows a screenshot of the CKAN Tag Manager.

\begin{figure}[ht]
\begin{center}
\includegraphics[width=\columnwidth]{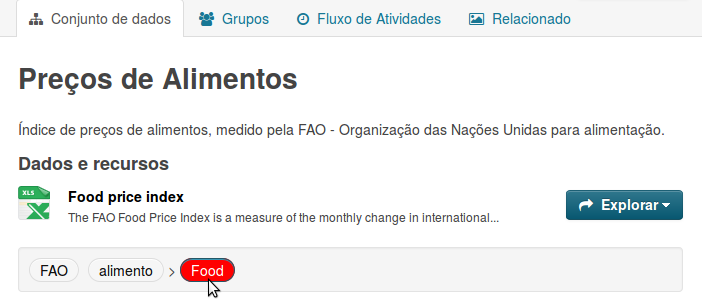}
\caption{Detail of dataset in an ODP. The dataset is tagged with two tags, and one of them (\texttt{alimentos}) is connected to the global tag \texttt{http://stodap.org/tags/Food} through the \texttt{muto:hasTag} property.}
\label{fig:local_link}
\end{center}
\end{figure}

The CKAN Semantic Tags plugin implements the connection between a CKAN instance and the Global Tag Server.
Each local tag can be associated to a global tag from the server.
After the association, datasets linked with a local tag also point to the global server, as shown in \autoref{fig:local_link}.

\begin{figure}[ht]
\begin{center}
\includegraphics[width=\columnwidth]{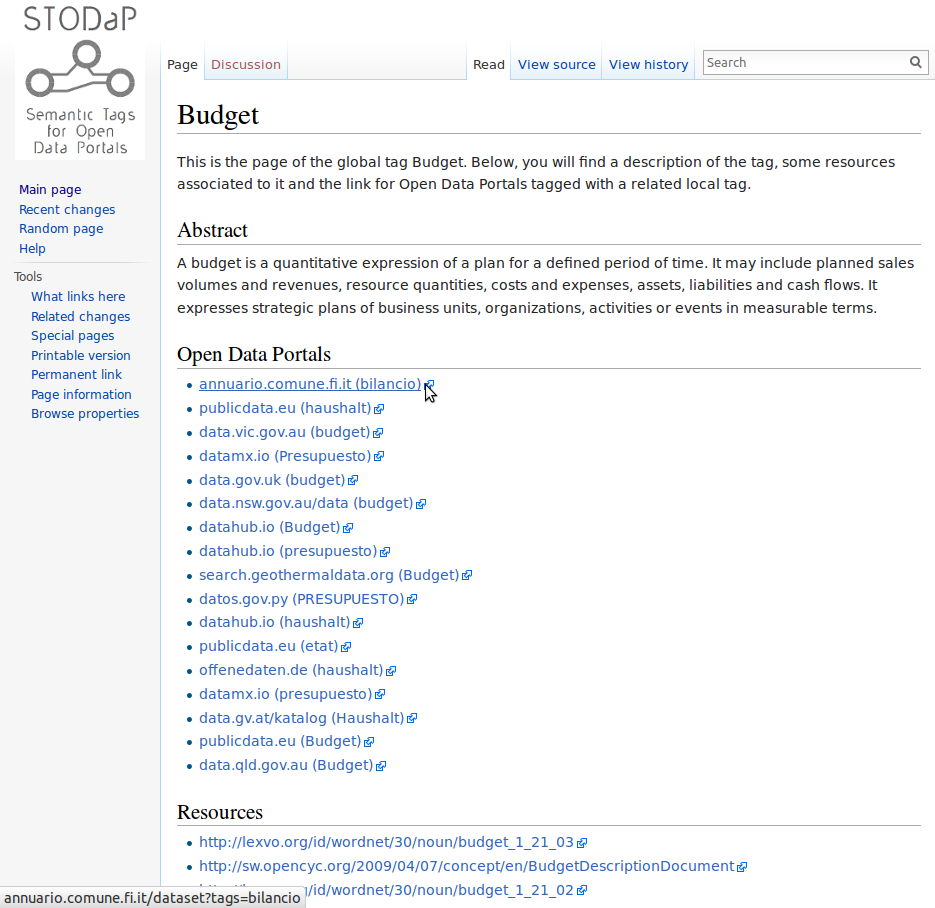}
\caption{Semantic Tag Server for Open Data Portals. Example of the global tag \texttt{Budget}, which is linked to 59 Open Data Portals and 10 semantic resources. In the screenshot, we illustrate a few of them.}
\label{fig:tag_server}
\end{center}
\end{figure}

The tag server is implemented using the collaborative \emph{MediaWiki}. 
Specially, the \emph{Semantic MediaWiki} extension~\cite{Kroetzsch2007} is used in order to include properties and integrate the global tags in the LOD Cloud, through the export of RDF files.
The page of a global tag is shown in \autoref{fig:tag_server}.
Each global tag page is build using semantic templates and forms\footnote{\url{https://www.mediawiki.org/wiki/Extension:Semantic_Forms}}, in order to facilitate consistency and coherency and to be more user-friendly.

\section{Results}
\label{sec:results}

We describe in this section some results achieved with the STODaP approach.
At the global level, it was possible to implement the global tags server and to test the performance.

\subsection{STODaP Server}

In order to test the system, an open-source implementation of STODaP was created and deployed at \url{http://stodap.org}.
The following approach was used create 100 global tags at the server:
\begin{itemize}
	\item From the 220,567 tags harvested, we selected the 200 that were used in more portals; 
	\item Using the Lexvo.org service, we found URI candidates to represent the tag meaning via the \texttt{lexvo:means} property;
	\item Using the Lexvo.org service, we found translations and synonyms for the tags via the \texttt{rdf:seeAlso} and \texttt{lexvo:translate} properties;
	\item We searched for the translations and synonyms in the harvested tags and included the results as \texttt{muto:taggedResources}, together with the portals tagged with the original term;
	\item Using the Semantic Measures Library\footnote{\url{http://www.semantic-measures-library.org}}, we searched for semantic similar global tags, which were added as \texttt{skos:related}.
\end{itemize}

\begin{figure}[tb]
\begin{center}
\includegraphics[width=\columnwidth]{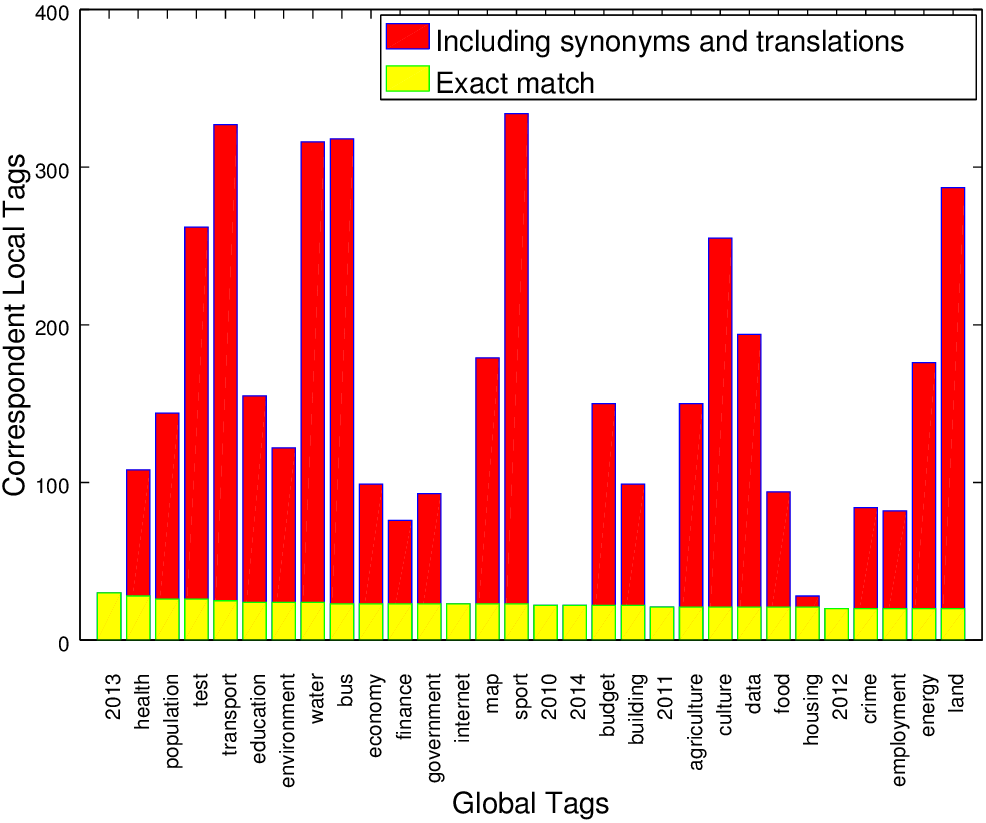}
\caption{Correspondence between local and global tags. The yellow bar shows the number of exact occurrences of the tag in ODPs. The red bar shows the improvement when considered translations and synonyms, which can also occur in a same portal. This explains the numbers over 90. }
\label{fig:results_tagged_resources}
\end{center}
\end{figure}

The occurrence of the original tags among the portals, and the results after including the translations and synonyms can be seen in \autoref{fig:results_tagged_resources}. 
The graphic shows the 30 most used tags, and the achieved increment in the number of relations. 
The occurrence of tags denoting years can also be noticed.
Obviously these tags have no synonyms nor translations, and thus no increment is shown. 
It is also worth mentioning that the tag \texttt{{test}} is the fourth most used one.
This fact is probably related to the early stage of development of some portals. 

\subsection{Local Level}

At the local level, the main potential achievements are at the tag curation process.
As shown in \autoref{fig:similarity}, a considerable number of pairs of tags differ only by capital or accented characters.
Using the naive approach to merge similar tags in every portal would result in reducing the number of 14,168 local tags, which represents 6.4\% of the total number of tags.
Lowercase and unaccented tags differing by a Levenshtein-distance from 0 to 2 represent a total of 35,066 pairs, or 15,8\% from the whole tag universe.
However, as discussed above, this approach can lead to false-positives and thus requires manual checking.

\section{Related work}
\label{sec:rel_works}

The vast majority of scientific works about tagging and semantics focus on a different kind of context in relation to ours.
Grubbers seminal paper~\cite{Grubber2007}, and others such as \cite{Halpin2007, Marchetti2007, Knerr2006, Mika2007, Kim2008, Kim2011} give interesting perspectives about the tagging activity and its relation to semantics, but always in the folksonomy (or collaborative tagging) context.
In this case, tags are attributed to resources by the crowd, and thus, their quality are enhanced by crowd-selection mechanisms.
This is applicable to platforms such as del.icio.us of flicker, where several users can tag the same resource.
However, in the open data context tags are only attributed by system managers.
The tag server approach described in this paper adds collaborative reconciliation layer over the ODPs.

In relation to the metrics for tagging environments, some related ideas could be found in the literature.
For example, \cite{Umbrich2015} presents a framework to evaluate the quality of ODPs. 
Among the applied quality metrics, three of them -- \emph{Usage}, \emph{Completeness} and \emph{Accuracy} -- are related to metadata keys, which tags are part of. 
\emph{Usage} establishes which metadata keys are actually used in a portal; \emph{Completeness} evaluates the presence of non empty values; and \emph{Accuracy} checks if metadata adequately describes the data.
However, this metric is not applied for tags.

Laniado and Mika did a similar analysis over hashtags on Twitter~\cite{Laniado2010}.
Their work is focused in answering if Twitter hashtags constitute \emph{strong identifiers} for the semantic web. 
To achieve this, four metrics are used: frequency of hashtags; specificity, which is the deviation from the use of them without being a hashtag; consistency; and stability over time.

The problem of semantic lifting in ODPs was tackled by~\cite{DBLP:conf/i-semantics/ErmilovAS13,Ding2011325}. 
In~\cite{Waal2014}, a strategy for lifting datasets in ODPs to the Linked Data cloud is presented. 
In all these works, however, the semantic lifting refers to the datasets, and not to metadata.

There also has been some work done with regard to metadata reconciliation~\cite{6406217,vanhooland_dh_2012}.
However, to the best of our knowledge none of them has been specifically applied to open data portals or leverages tag curation as proposed by STODaP.


\section{Conclusions}
\label{sec:conclusions}

In this paper, we presented an approach for metadata reconciliation among Open Data Portals.
The use of tags was analysed, and several problems were found, such as a low tag reuse rate and the overall existence of different tags for the same meaning.
On the analysis we also found that several portals share the same tags, showing that tags have a good potential to be linking elements among datasets.
Converting tags into semantic identifiers was also shown as a viable option, even though more sophisticated methods have to be investigated. 
Based on these findings, we derived the STODaP approach, which comprises two parts: 
a local one, aimed at cleaning up and enhancing the quality of ODPs tags, and 
a global one, for connecting ODPs through semantic tags.
The implementation of both shows that significant enhancements can be achieved both at the individual ODPs and the global levels.

Future research and development includes a tag suggestion approach for ODPs which takes into account the related tags at the tag server, using collective knowledge as in~\cite{Sigurbjornsson2008}.
Using the possibly structured data of the ODPs in order to improve tagging suggestions is also a research direction that should be followed.
At the global level, an interesting approach is to detect the emergence of schemas from the tags, as described in~\cite{Robu2009}.
We will also call for the attention of the open data community in order further to advance collaborative strategies for enriching the tag server.
For STOdAP to realize its full potential, ODP administrators and users should be involved and (meta)data literacy needs to be improved.


\section*{Acknowledgements}

This work was supported by a grant of the European Commission within the Horizon2020 framework programme for the project OpenBudgets.
A. Tygel is supported by CAPES/PDSE grant 99999.008268/2014-02. 
M. L. M. Campos is partially supported by CNPq--Brazil.

\bibliographystyle{abbrv}
\bibliography{IEEEabrv,bibliography}  

\end{document}